\documentclass[twocolumn,floatfix,secnumarabic,tightenlines,amssymb,amsmath,aps,prl]{revtex4}

\usepackage{graphicx}
\usepackage{bm}

\begin{document}
\preprint{VERSION 1.0}

\title{Diameter and Chirality Dependence of Exciton Properties in Carbon Nanotubes}

\author{Rodrigo B. Capaz$^{1,2,5,6}$\email{capaz@if.ufrj.br}}
\author{Catalin D. Spataru$^{3,5,6}$}\author{Sohrab Ismail-Beigi$^4$}\author{Steven G. Louie$^{5,6}$}

\affiliation{
$^1$Instituto de F\'isica, Universidade Federal do Rio de Janeiro, Caixa Postal 68528, Rio de Janeiro, RJ 21941-972, Brazil\\
$^2$Divis\~ao de Metrologia de Materiais, Instituto Nacional de Metrologia, Normaliza\c c\~ao e Qualidade Industrial - Inmetro,  R. Nossa Senhora das Gra\c cas 50, Xer\'em, Duque de Caxias, RJ 25245-020, Brazil\\
$^3$Center for Integrated Science and Engineering and Center for Electron
Transport in Molecular Nanostructures, Columbia University, New York, NY 10027, USA \\
$^4$Department of Applied Physics, Yale University, New Haven, Connecticut 06520, USA\\
$^5$Department of Physics, University of California at Berkeley, Berkeley, CA 94720, USA\\
$^6$Materials Science Division, Lawrence Berkeley National Laboratory, Berkeley, CA 94720, USA }

\begin{abstract}
We calculate the diameter and chirality dependences of the binding energies, sizes, 
and bright-dark splittings of excitons in semiconducting single-wall carbon nanotubes (SWNTs). Using results and 
insights from {\it ab initio} calculations, we employ a symmetry-based, variational method based on 
the effective-mass and envelope-function approximations using tight-binding 
wavefunctions. Binding energies and spatial extents show a leading dependence 
with diameter as $1/d$ and $d$, respectively, with chirality corrections providing 
a spread of roughly 20\% with a strong family behavior. Bright-dark exciton 
splittings show a $1/d^2$ leading dependence. We provide analytical expressions 
for the binding energies, sizes, and splittings that should be useful to guide future 
experiments. 
\end{abstract}

\date{\today}

\maketitle

Diameter and chirality trends are one of the most useful concepts in nanotube 
science. Often, new physics arises when the diameter and chirality dependences 
of a given property are fully disclosed. A classic example is the 
analysis of ``family patterns'' in  optical transitions combined with the
diameter dependence of vibrational frequencies that paved the way to reliable
$(n,m)$ assignments of single-wall carbon nanotubes (SWNTs) \cite{jorio013} and posed
the fundamental  ``ratio problem'' \cite{bachilo02}. Therefore, a reliable 
determination of diameter and chirality trends of a given
nanotube property, even when this is accomplished by simplified models,  
is often as important as determining accurately that property for a limited 
number of tubes. Moreover, when a reliable model for trends is coupled with
an accurate {\it ab initio} theory that determines its parameters, the model
acquires quantitative and predictive powers. 

The exciton concept solved the ``ratio problem'' \cite{spataru041}, and it is 
now widely accepted that the optical spectra of carbon nanotubes are dominated 
by exciton features \cite{spataru041,ando97,chang04,perebeinos04,zhao04}. 
Recent experiments based on two-photon spectroscopy \cite{wang05,dukovic05,maultzsch05} and 
Raman spectroscopy on 
electrochemically doped samples \cite{wang06} have provided the first experimental 
evaluations of exciton binding energies for a few single-wall carbon nanotubes 
(SWNTs). However, a full description of diameter and chirality dependences of 
exciton properties in SWNTs has not yet been provided, neither experimentally nor 
theoretically. {\it Ab initio} calculations are restricted to a few small-diameter 
tubes \cite{spataru041, chang04}. Perebeinos {\it et al.} \cite{perebeinos04} have 
extracted scaling relations of binding energies and sizes with diameter from model 
calculations, but the chirality dependence has not been addressed. Semi-empirical 
calculations have also been done for a larger variety of tubes \cite{zhao04}, but 
again systematic diameter and chirality trends have not been extracted. Finally,
the important issue of bright-dark exciton splittings have been addressed in detail 
by considerably fewer calculations \cite{spataru05, perebeinos051, chang06}. 

In this work, we calculate the full diameter and chirality dependences of exciton 
properties in SWNTs. We employ a symmetry-based, variational, tight-binding method, 
based on the effective mass and envelope function approximations \cite{knox,brown87}. 
Since we explictly impose the symmetry of the exciton wavefunction, 
we can calculate properties of bright and dark excitons. Our model is
parametrized by {\it ab initio} results. 
We calculate binding energies and sizes for
the lowest-energy bright excitons (those usually associated with the E$_{11}$
singularity in the single particle joint density of states), as well as dark-bright
exciton splittings for a large number of SWNTs. From these results, we extract
reliable analytical expressions for the diameter and chirality dependences of
such properties.

Our variational exciton wavefunction is written as:
\begin{equation}
\psi(\vec{r}_e,\vec{r}_h) = C\sum_{v,c}A_{vc}\phi_{c}(\vec{r}_e)\phi_{v}^*(\vec{r}_h)e^{\frac{(z_e-z_h)^2}{2\sigma^2}},\label{exc:psi}
\end{equation}
where $\phi_{c}(\vec{r}_e)$ and $\phi_{v}(\vec{r}_h)$ are conduction (electron) and 
valence (hole) single-particle states. The sum is restricted to the four band-edge states 
$c=\pm m$ and $v=\pm m$ from the top of the valence and bottom of the conduction 
bands. Note that both valence and conduction band edges are 2-fold degenerate for 
both zigzag and chiral tubes, once time-reversal symmetry is considered 
\cite{vukovic02,barros06}. The single-particle wavefunctions are labeled by their 
quasi-angular momentum quantum numbers $+m$ and $-m$ and they are taken from 
properly symmetrized wavefunctions of graphene expanded in a $\pi$-orbital 
tight-binding basis. \cite{vukovic02} The coefficients $A_{vc}$, responsible for the 
quantum interference between pair excitations, are then completely determined by 
symmetry, as described in Table \ref{tab:1}.
\begin{table}
\begin{footnotesize}
\caption{Symmetries, degeneracies, optical activities and coefficients $A_{vc}$ for 
excitons in zigzag and chiral tubes. The symmetries are described by the irreducible 
representations in both group of the wave vector \cite{barros06} and line group 
\cite{vukovic02}(in parenthesis) notations. The label $m'$ is the quasi-angular momentum quantum 
number of the double-degenerate exciton.  \label{tab:1}}

\begin{tabular}{ccccccc}\hline
& & Zigzag \\
\hline
Symmetry & Deg. & Activity & $A_{++}$ & $A_{--}$ & $A_{+-}$ & $A_{-+}$ \\
\hline
$A_{1u}(_0B_0^-)$ & 1 & dark &  1 & -1  & 0 & 0 \\
$A_{2u}(_0A_0^-)$ & 1 & bright & 1  & 1 & 0 & 0 \\
$E_{m',u}(_0E_{m'}^-)$ & 2 & dark & 0 & 0 & $\pm $1  & $\mp $1  \\
\hline
& & Chiral \\
\hline
Symmetry & Deg. & Activity & $A_{++}$ & $A_{--}$ & $A_{+-}$ & $A_{-+}$ \\
\hline
$A_1(_0A_0^+)$ & 1 & dark &  1 & 1 & 0 & 0 \\
$A_2(_0A_0^-)$ & 1 & bright & 1 & -1  & 0 & 0 \\
$\mathbb{E}_{m'}(k) + \mathbb{E}_{-m'}(-k) (_kE_{m'})$ & 2 & dark & 0 & 0 & $\pm $1  & $\mp $1  \\

\end{tabular}
\end{footnotesize}
\end{table}

\begin{figure}
\includegraphics[width=7.5cm]{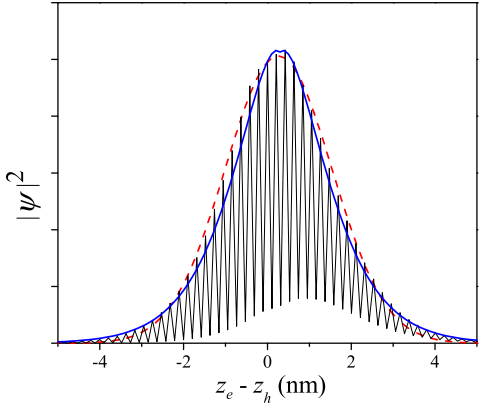}
\caption{(Color online) Lowest-energy singlet exciton wavefunction of the (11,0)
tube. Black thin line: {\it Ab initio} $|\psi|^2$ after integrating out on the 
coordinates perpendicular to the tube. 
Red thick dashed line: Envelope fit using a gaussian. 
Blue thick line: Envelope fit using a Whittaker function. Notice that the two fits
are almost indistinguishable.} \label{fig:gaussian}
\end{figure}

We choose a gaussian envelope function. This choice is justified both by a fit of 
the {\it ab initio} exciton wavefunctions \cite{spataru041}, as shown in Fig. \ref{fig:gaussian}, and by an 
analogy with the regularized Coulomb potential problem in 1D 
\cite{loudon59,banyai87}, for which the ground-state Whittaker function 
closely resembles a gaussian. The gaussian width or exciton size $\sigma$ is the 
only variational parameter in the problem. The constant $C$ normalizes the 
exciton wavefunction:
$\int\int |\psi(\vec{r}_e,\vec{r}_h)|^2 d\vec{r}_e d\vec{r}_h = 1$.

We minimize the exciton energy that is composed of three terms: direct, exchange and 
kinetic energies. Here, we treat singlet excitons only. The direct term is written as:
\begin{eqnarray}
<K^d> =\int \psi^*(\vec{r}_e,\vec{r}_h) V_C^{scr}(\vec{r}_e-\vec{r}_h) \psi(\vec{r}_e,\vec{r}_h) d\vec{r}_e d\vec{r}_h \nonumber \\
=C^2\sum_{vc,v'c'}A_{vc}^*A_{v'c'}\sum_{\vec{R}_1,\vec{R}_2}c_v(\vec{R}_1)c_{v'}^*(\vec{R}_1)c_c^*(\vec{R}_2)c_{c'}(\vec{R}_2)\times \nonumber \\
\times e^{-\frac{(Z_1-Z_2)^2}{\sigma^2}}U_{Ohno}^{scr}(|\vec{R_1}-\vec{R_2}|),
\label{eq:dir}
\end{eqnarray}
where $V_C^{scr}$ is the screened Coulomb interaction and we wrote the direct energy in terms of the tight-binding expansion coefficients of the single-particle wavefunctions in a $p_z$-orbital basis $\varphi (\vec r - \vec R_i)$ centered in the atomic 
positions $\vec R_i$: 
\begin{equation}
\phi_n(\vec r)=\sum_i c_n(\vec R_i) \varphi (\vec r - \vec R_i).
\end{equation}
The Coulomb integrals between sites are parametrized by the Ohno formula \cite{ohno64}:
\begin{equation}
U_{Ohno}^{scr}(R)=\frac{U_0}{\epsilon\sqrt{(\frac{4\pi\epsilon_0}{e^2}U_0R)^2 + 1}}.
\label{eq:ohno}
\end{equation}
The onsite Coulomb repulsion $U_0=16$ eV and the dielectric constant $\epsilon=1.846$ 
are chosen to reproduce the {\it ab initio} values for the binding energy and 
bright-dark exciton splittings for the (11,0) tube and kept constant for all other 
tubes. The exchange energy is given by:
\begin{eqnarray}
<K^{x}> = 2 \int \psi^*(\vec{r}_e,\vec{r}_e) V_C(\vec{r}_e-\vec{r}_h) \psi(\vec{r}_h,\vec{r}_h) d\vec{r}_e d\vec{r}_h  \nonumber \\
=2C^2\sum_{vc,v'c'}A_{vc}^*A_{v'c'}\sum_{\vec{R}_1,\vec{R}_2}c_v(\vec{R}_1)c_c(\vec{R}_1)c_{v'}^*(\vec{R}_2)c_{c'}^*(\vec{R}_2)\times \nonumber \\
\times U_{Ohno}(|\vec{R}_1-\vec{R}_2|). \label{eq:exc}
\end{eqnarray}
In this case, the unscreened Coulomb interaction $V_C$ is parametrized by taking 
$\epsilon=1$ in Eq.(\ref{eq:ohno}). Finally, the kinetic energy associated with the 
exciton relative coordinate is simply that of a gaussian envelope:
\begin{equation}
<T>=\frac{\hbar^2}{4m^*\sigma^2}, \label{eq:kinetic}
\end{equation}
where the exciton reduced mass $m^*$ is given by $1/m^* = 1/m_e + 1/m_h$. We use 
the diameter- and chirality-dependent electron ($m_e$) and hole ($m_h$) effective 
masses obtained from tight-binding calculations \cite{jorio05}.
	
To test our model, we compare in Table \ref{tab:2} our variational binding energies 
with {\it ab initio} ones obtained from solving the Bethe-Salpeter equation \cite{spataru041}
for a few zigzag tubes. The agreement is excellent, 
except for the E$_{22}$ exciton in the (7,0) SWNT. This discrepancy can be 
understood: For the small-diameter 
(7,0) tube, the E$_{22}$ exciton size becomes extremely small ($\sigma = 5.3$ \AA) and 
therefore the envelope-function approximation is not expected to be valid in this 
regime. Notice also that our model correctly captures the family oscillations in 
the binding energy.

\begin{table}
\begin{footnotesize}
\caption{{\it Ab initio} and model binding energies for bright E$_{11}$ and E$_{22}$ 
excitons for a few small-diameter SWNTs. \label{tab:2}}

\begin{tabular}{c|cc|cc}\hline
&\multicolumn{2}{c|}{$E_b^{11}$ (eV)} & \multicolumn{2}{c}{$E_b^{22}$ (eV)}      \\
\hline
Tube & {\it Ab Initio} & Model & {\it Ab Initio} & Model  \\
(7,0) & 0.89 & 0.87 & 1.13 & 1.61 \\
(8,0) & 0.99 & 1.03 & 0.86 & 0.92 \\
(10,0) & 0.76 & 0.68 & 0.95 & 1.09 \\
(11,0) & 0.76 & 0.76 (fitted) & 0.72 & 0.75 \\
\end{tabular}
\end{footnotesize}
\end{table}

\begin{figure}
\includegraphics[width=7.5cm]{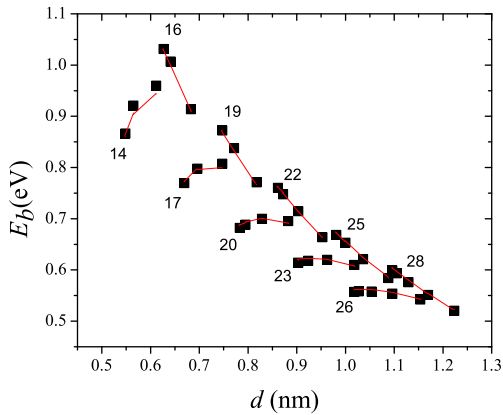}
\caption{Binding energies for the lowest-energy bright excitons in 38 SWNTs with varying diameter
and chirality. The dots are our model results and the red lines represent the analytical
fit using Eq.(\ref{eq:analyt}). The labels indicate the $(2n+m)$ families.} \label{fig:binding}
\end{figure}

Fig. \ref{fig:binding} shows the binding energy of the E$_{11}$ bright exciton as a function 
of diameter for 38 SWNTs covering the full range of chiralities. The $(2n+m)$ family 
indices are indicated in the figure. One clearly sees the well-known family pattern 
reminiscent from the so-called Kataura plots for optical transition energies. \cite{jorio05} 
As expected, binding energies decrease with tube diameter. Chirality effects are also 
strong, contributing to about 20\% spread in the binding energies for a given 
diameter. It is clear that excitons in $(2n+m)$ mod $3=1$ (MOD1) tubes have generally 
larger binding energies than in $(2n+m)$ mod $3=2$ (MOD2) tubes. 

\begin{figure}
\includegraphics[width=7.5cm]{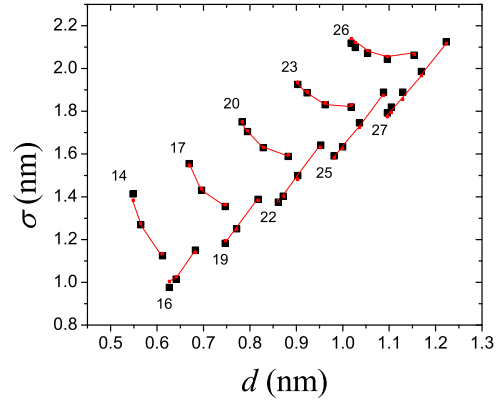}
\caption{Sizes of the bright lowest-energy excitons for 38 SWNTs with varying diameter
and chirality. The dots are our model results and the red lines represent the analytical
fit using Eq.(\ref{eq:analyt}). The labels indicate the $(2n+m)$ families.} \label{fig:size}
\end{figure}

Fig. \ref{fig:size} 
shows the exciton sizes as a function of diameter. Again, as expected, exciton sizes 
increase with diameter and they show the opposite MOD1-MOD2 trends as compared to 
binding energies. Notice that even for tubes as small as 0.5 nm in diameter the 
E$_{11}$ exciton sizes are already several times larger than the carbon-carbon bond, 
thus justifying the use of the envelope-function approximation. 

Analytical expressions for diameter and chirality dependences, although sometimes 
lacking a deeper physical justification, can be extremely useful for a quick 
evaluation of a variety of nanotube properties. We succeeded in finding simple yet 
very accurate analytical approximations for both binding energies and sizes:
\begin{center}
\begin{eqnarray}
E_b=\frac{1}{d}\left(A+\frac{B}{d}+C\xi+D\xi^2\right) \nonumber \\
\sigma=d(E+F\xi+G\xi^2), \label{eq:analyt}
\end{eqnarray}
\end{center}
where $d$ is the tube diameter in nm and $\xi=(-1)^{\nu}cos3\theta/d$ captures the 
chirality dependence \cite{capaz05}. The best fits are given by $A=0.6724$ eV.nm, 
$B=-4.910\times10^{-2}$ eV.nm$^2$, $C=4.577\times 10^{-2}$ eV.nm$^2$, 
$D=-8.325\times 10^{-3}$ eV.nm$^3$, $E=1.769$, $F=-2.490\times 10^{-1}$ nm and 
$G=9.130\times10^{-2}$ nm$^2$. These analytical fits are plotted in red lines in Figs. \ref{fig:binding} and \ref{fig:size}, together with the numerical results. The agreement is nearly perfect.

Our theory also allows for an estimation of chirality and diameter dependences of 
exciton splittings among exciton states of the ground-state complex of the
same E$_{ii}$. These splittings are fundamental to understand a 
variety of optical properties of carbon nanotubes, such as the quantum efficiency 
for light emission and the exciton radiative lifetime \cite{spataru05,perebeinos051}. 
{\it We find that the lowest-energy exciton for all SWNTs is the singly-degenerate 
dark state}, due to its vanishing exchange energy \cite{spataru05}. 
Defining the exciton 
splittings from the lowest-energy exciton to the bright exciton and to the 
double-degenerate dark exciton as $\delta_1$ and $\delta_2$, we find the following 
dependence on diameter and chirality:
\begin{equation}
\delta_i=\frac{1}{d^2}\left(A_i+B_i\xi+C_id\xi^2\right), \label{eq:delta}
\end{equation}
with $A_1 = 18.425$ meV.nm$^2$, $B_1=12.481$ meV.nm$^3$, $C_1=-0.715$ meV.nm$^3$, $A_2=32.332$ meV.nm$^2$, $B_2=7.465$ meV.nm$^3$ and $C_2=-2.576$ meV.nm$^3$. So, in disagreement with Perebeinos {\it et al.} \cite{perebeinos051}, we find 
the leading dependence of bright-dark splittings to be $1/d^2$. This is precisely the dependence of the exchange energy
$<K^x>$ on diameter.

It is instructive to explain on physical grounds the leading dependences on diameter of the exciton sizes, binding
energies and the bright-dark splittings. The exciton sizes $\sigma$ scale like
$d$ because the 1D Coulomb potential is smoothed out or regularized over the scale of the tube 
diameter $d$ and this sets the length scale of the bound state (recall that in a pure 1D system 
with no lateral size, the Coulomb potential gives a delta-function ground state with infinite
binding energy). The binding energies go like $1/d$ because $\sigma$ scales like $d$ and 
Coulomb interactions go like inverse distance \cite{perebeinos04}. The scaling of dark-bright 
splittings mirrors the scaling of the exchange energy $<K^x>$ which goes like $1/d^2$ 
because $<K^x>$ is the self-interaction of a neutral charge distribution  
with dipole moments: The long-range part (from distances larger than $d$) can
be written as $\int_d^\infty  dx / x^3  \sim 1/d^2$.

We now compare our results to the available experimental determinations of the 
exciton binding energies to date. Two-photon spectroscopy have been performed for SWNTs in a polymeric matrix \cite{wang05,dukovic05} and in D$_2$O solution wrapped by a surfactant \cite{maultzsch05}. 
These environments should provide extra screening, so these results should not be directly compared with {\it ab initio} theory for isolated tubes. However, in our variational scheme, it is very easy to investigate the influence of screening and to adjust the dielectric constant $\epsilon$ 
to match the experimental results. In fact, we find that binding energies follow 
very nicely the scaling $E_b \propto \epsilon^{-1.4}$ proposed by Perebeinos 
{\it et al.} \cite{perebeinos04}. Therefore, it is straightforward to apply 
Eq.(\ref{eq:analyt}) for SWNTs in {\it any} environment, provided that one scales 
the binding energies by using the appropriate phenomenological dielectric constant. 
For instance, taking $\epsilon = 3.049$ gives binding energies in excellent agreement (standard deviation of 0.02 eV) for all 13 SWNTs measured by Dukovic {\it et al.} \cite{dukovic05}. Similarly, the results of Maultzsch {\it et al.} \cite{maultzsch05} for 6 different SWNTs are reproduced with a standard deviation of 0.03 eV using a slightly larger dielectric constant $\epsilon = 3.208$.

In another recent experiment, Raman spectroscopy under electrochemical doping was 
used in nanotubes coated with a surfactant to give 0.62 eV and 0.49 eV for the 
binding energies of excitons associated with E$_{22}$ transitions in the (7,5) 
and (10,3) SWNTs, respectively. \cite{wang06} 
We have also calculated the binding energies of 
those excitons. It should be noted that for E$_{22}$ excitons, 
the MOD1-MOD2 oscillations in the binding energies are inverted, i.e., MOD2 tubes 
have larger binding energies than MOD1 tubes of similar diameter. In fact, in
discrepancy with experiment, we find 
that (7,5) and (10,3) tubes should have E$_{22}$ excitons with similar binding 
energies, even though the latter has a larger diameter. By using $\epsilon = 2.559$ 
we find the best possible ``average'' agreement with experiment: 0.54 eV for the (7,5) and 0.55 eV for the (10,3) nanotube.    

In conclusion, we have determined the full diameter and chirality dependence of
exciton binding energies, sizes, and splittings in semiconducting SWNTs. 
All these exciton properties have strong diameter and chirality dependences,
with a distinct family behavior. Comparisons between theoretical and experimental
binding energies should be exercised with care, by ackowledging environmental 
screening effects.  Our results should provide an useful guide to the interpretation 
of recent and future experimental determinations of exciton binding energies and
other properties.

We acknowledge useful discussions with A. Jorio and T. G. Rappoport. RBC acknowledges 
financial support from the John Simon Guggenheim Memorial Foundation and Brazilian 
funding agencies CNPq, FAPERJ, Instituto de Nanoci{\^e}ncias, FUJB-UFRJ and MCT. 
Work partially supported by NSF Grant No. DMR04-39768 and DOE Contract No. DE-AC02-05CH11231.   

\bibliography{var-condmat.bib}

\end{document}